# Spin disorder competing with positional symmetry breaking governs the metal-insulator behavior in oxide paramagnets

Jia-Xin Xiong[1,†], Xiuwen Zhang[1], and Alex Zunger[1,*]

[1]Renewable and Sustainable Energy Institute, University of Colorado, Boulder, Colorado 80309, USA

## Abstract

Numerous transition-metal oxides have low-temperature, long-range-ordered antiferromagnetic (AFM) states that are generally insulating, and high-temperature, disordered paramagnetic (PM) phases. The latter can be either insulating (predicted here for $NaFeO_3$), or metallic (predicted here and previously observed in $NaOsO_3$). Similar distinctions have been traditionally affected in strongly correlated models by the value used for Coulomb repulsion $U$. Here we show an alternative, strong-correlation-free ($U$=0) view suggesting that the distinction between insulating and metallic PM phases is governed by the competition between local magnetic moment disorder and the polymorphous distribution of off-center atomic displacements. Such parameter-free, energy-lowering symmetry breaking density functional calculations provide a framework for understanding metal-insulator behaviors across different quantum materials in terms of measurable local structural and magnetic parameters.

† Corresponding author: Jia-Xin.Xiong@colorado.edu
* Corresponding author: alex.zunger@colorado.edu

Magnetic transition-metal oxides generally have a long-range ordered antiferromagnetic (AFM) ground phase, followed above the Néel temperature by a substitutionally disordered paramagnetic (PM) phase. Table I illustrates oxide compounds that are insulating as AFM and insulating as PM phases (already just above the Néel temperature, Group I). Recently, a *metallic* $ABO_3$ paramagnet $NaOsO_3$, (Group II) was shown to evolve from an insulating AFM phase just above the Néel temperature [1,2]. This is a rather unusual case as other metallic paramagnets appear only far above the Néel temperature, presumably due to extensive disorder ($YNiO_3$ [3]). In cases where the PM phase is metallic, the ordered AFM ground state phase is often not observed at all ("persistent metals" such as $SrVO_3$ [4] and $LaNiO_3$ [5,6]), or the phase from which the PM evolves is instead ferromagnetic ($YTiO_3$ [7]) or nonmagnetic ($VO_2$ [8,9]). The PM metals discussed here are special, called "gapped metals", and different than the traditional element metals, as the former have an *internal energy gap* between their conduction band minimum (CBM) and valence band maximum (VBM), yet the Fermi level resides either above CBM as in $SrVO_3$ [4], or below VBM, as in $LaNiO_3$ [6] and PM $NaOsO_3$. The question addressed in this paper is what decides if the PM phase is insulating (as in Group I) or metallic (as in Group II). Here we provide a predictive framework that identifies the microscopic competition between magnetic and positional symmetry breaking that determines the metallic vs insulating character of the PM phase.

Table I. Transition-metal compounds classified by insulating or metallic behaviors in their antiferromagnetic (AFM) and paramagnetic (PM) phases. Short abbreviations for terminologies: "Polar." for polarization; "Orth." for orthorhombic, "Mono." for monoclinic; "JT" for Jahn-Teller distortion, "disprop." for disproportionation, "Mag." for magnetic. The DFT-calculated band gaps for insulating phases are included. The band gaps are calculated by SCAN.

| Compound | Crystal structure | Symmetry breaking type | Metal or insulator as AFM (gap) | Metal or insulator as PM (gap) |
|---|---|---|---|---|
| | | Group I | | |
| $LaFeO_3$ | Orth. | Spin polar. | Insulator (1.49 eV) | Insulator (0.73 eV) |
| $LaTiO_3$ | Orth. | Spin polar. | Insulator (0.10 eV) | Insulator (0.13 eV) |
| $YNiO_3$ | Mono. | Mag. disprop. | Insulator (0.93 eV) | Insulator (0.22 eV) |
| $LaMnO_3$ | Orth. | Spin polar.; JT | Insulator (0.34 eV) | Insulator (0.05 eV) |
| $NaFeO_3$ | Orth. | Spin polar. | Insulator (0.17 eV) | Insulator (0.29 eV) |
| | | Group II | | |
| $NaOsO_3$ | Orth. | Spin polar. | Insulator (0.22 eV) | Metal |

*Previous explanations:* There are a few factors, discussed in the literature, that might make the PM phase metallic: In 1951, Slater [10] suggested that the existence of gapped (insulating) antiferromagnets, such as NiO and MnO is a consequence of cell-doubling $(A\text{-}\uparrow)_n/(B\text{-}\downarrow)_n$ repeated in a long-range order (LRO) fashion, and leading to band repulsion forming the AFM crystal band gap. As a consequence of Slater's reliance on LRO to explain AFM gapping, it followed that the *absence* of LRO in the PM phase would deprive the system from band repulsion, thus leading to universal stabilization of a *metallic* PM. On the other hand, simplified band structure approaches [11–14] automatically predicted universal metallic PM phases when replacing the local moments by the compensated global average zero moment rather than allowing for local moments to be nonzero and compensating just on average to zero.

*Approach:* Rather than use a theoretical approach that assumes a uniform magnetic moment configuration for the PM phase (represented by a minimal-size repeated unit cell), here we use a supercell approach of spin special-quasirandom structure (spin-SQS) that permits a distribution of local moments as well as distribution of atomic displacements. This can predict either metallic or insulating PM phases—depending on the specific electronic and magnetic descriptors of the specific compounds. Such predictive theory [15] describes the PM phase as a distribution of nonzero local magnetic moments on each atom, leading to global spin compensation of the phase. The PM structures are optimized by relaxing cell-internal atoms in fixed lattice constants as obtained from the AFM structures. The spin-SQS method [15] represents a configurational average of many snapshots rather than a single snapshot, and includes the local atomic environment, hence describing realistically the spin-disordered PM phase without LRO. Total energy minimization of the supercell in search of symmetry breaking shows in the PM phase spin-disordered local magnetic moments with non-zero atomic displacements, called "polymorphous network" configuration [16–21]. When applied to an electronic structure solver (Schrödinger equation), this symmetry-broken supercell model can accommodate, either a metallic or an insulating response, depending on the local positional and spin descriptors determined via energy lowering noted above. Applications to $LaFeO_3$ [6,22], NiO [16,23], MnO [16,23], $NbO_2$ [6], etc. (all corresponding to Group I cases) were previously shown to be all insulating PM phases. Here we specifically confront a system that can have a legitimate metallic ($NaOsO_3$) or insulating ($NaFeO_3$) PM phase, examining if it can be predicted from non-empirical first principles.

*Technical details:* We perform first-principles calculations using a pseudopotential density functional theory (DFT) method, including spin polarization for both AFM and PM phases, without involving strong correlation by excluding the on-site potential term *U* (which represents self-interaction reduction). We use the exchange correlation functional "strongly constrained and appropriately normed" (SCAN) [24]. For AFM phases, we use the primitive cell that allows spin-ordered AFM configurations. The AFM structures are obtained by relaxing both lattice constants and cell-internal atoms. For PM phases, instead of primitive cells, we use a large spin-SQS supercell of 160 atoms that allows positional and spin-disordered symmetry breaking. Spin-SQS uses the collinear magnetic calculations to simulate the spin disordered PM configurations and successfully provides the metallic PM phase of $NaOsO_3$. The collinear spin-SQS employed here represents a configurational limit of static disordered local moments and does not attempt to reproduce spin dynamics. Importantly, the quantity central to the present work is the distribution of local moment magnitudes and its electronic structure consequences, rather than the dynamical orientation of magnetic moments. We did DFT calculations also including spin-orbit coupling (SOC) and found that SOC has a minor effect on the band gap for $NaOsO_3$, consistent with Ref. [25]. More methodological details are given in Section A of Supplemental Material.

*Group I* compounds: This group including $LaTiO_3$ and $LaFeO_3$ corresponds to the chemical type $A^{3+}B^{3+}(O^{2-})_3$ where the ionic charge is $B^{3+}$. For $LaTiO_3$ this corresponds to $B=Ti^{3+}$ to an extra *d*-electron $d^1$ relative to the neutral B atom ($d^2s^2$). In the absence of symmetry breaking, the extra *d*-electron partially fills the broad conduction band and makes it a mean-field "false metal", which is inconsistent with experiments. Magnetic and positional symmetry breaking produce an insulating paramagnet (as shown in Figure S2(c) in Ref. [6]) for compounds of the $A^{3+}B^{3+}(O^{2-})_3$ group, with *occupied* isolated split-off flat bands that originate from the extra *d*-electron. On the other hand, for $LaFeO_3$ with $Fe^{3+}$, one finds a $d^5$ configuration. Before symmetry breaking, this creates a mean-field false metal (partially occupying the *principal valence band*). After symmetry breaking, this shows an insulator with *unoccupied* split-off flat bands in the gap. Calculations for these compounds give AFM insulators (gaps of 0.10 and 1.49 eV for $LaTiO_3$ and $LaFeO_3$, respectively) and PM insulators (gaps of 0.13 and 0.73 eV for $LaTiO_3$ and $LaFeO_3$, respectively).

*Group II compounds:* Shi et al. [1] and Vecchio et al. [2] synthesized and measured a compound that belongs to another type $A^{1+}B^{5+}(O^{2-})_3$ such as $Na^{1+}Os^{5+}(O^{2-})_3$, where both the AFM and the PM phases have the same orthorhombic crystallographic structures with the space group of *Pnma*. The ionic charge on the B-site is $B^{5+}$, relative to charge neutral Os atom ($d^6s^2$); this gives a *d*-orbital occupation $d^3$. We find in symmetry-broken DFT calculations that the AFM phase is insulating with the gap of 0.22 eV, while its PM phase is metallic, as observed in experiments [1,2]. Moreover, neutron and X-ray scattering experiments show that $NaOsO_3$ undergoes a continuous transition from the AFM insulator to PM metal near the Néel temperature ~410 K while retaining the average *Pnma* structure [26]. Subsequent inelastic X-ray scattering (RIXS) measurements further revealed magnetic excitations and short-range spin correlations persisting into the PM metallic state [27]. This group is related to Group I, in which both groups have an insulating AFM phase, but Group II has a metallic PM phase.

*Comparison with temperature-dependent crystallographic measurements:* The calculated structure of $NaOsO_3$ agrees closely with the neutron and X-ray diffraction measurements across the Néel temperature [26]. For the AFM phase, the calculated and experimental mean Os-O bond lengths are $1.95 \pm 0.00$ Å and $1.94 \pm 0.00$ Å respectively, while the mean Os-Os atomic distances are $3.77 \pm 0.00$ Å and $3.80 \pm 0.00$ Å experimentally. For the PM phase, the corresponding calculated values are $1.95 \pm 0.02$ Å and $3.77 \pm 0.01$ Å, compared with experimental values of $1.94 \pm 0.00$ Å and $3.80 \pm 0.00$ Å, respectively. The O-Os-O bond angles are also closely reproduced, whereas the Os-O-Os bond angles are slightly underestimated ~3°. This slight underestimation is due to the DFT-relaxed lattice constants that are less than 0.8% smaller than the experimental ones. Importantly, the conventional diffraction refinement [26] reports the average *Pnma* crystallographic structure, whereas the symmetry-broken spin-SQS relaxation resolves distributions of locally inequivalent environments, which broaden strongly in the PM phase despite the persistence of the average *Pnma* structure experimentally (see Figure 1). Details of compared parameters are given in Section B of Supplemental Material.

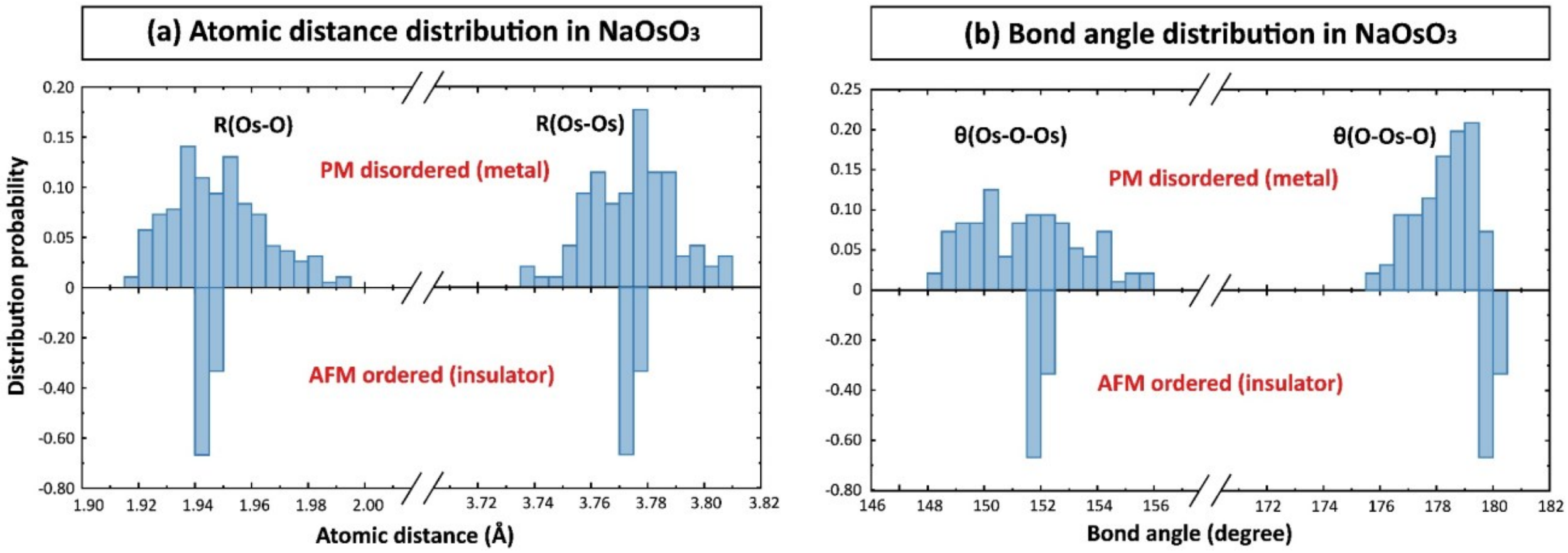

Figure 1. Symmetry-broken DFT-relaxed structures of PM and AFM phases of orthorhombic $NaOsO_3$: Distribution of (a) atomic distances including Os-O and Os-Os and (b) bond angles of Os-O-Os and O-Os-O. The bin widths of histograms in (a) and (b) are set to be 0.005 Å and 0.5 degree, respectively. The distribution probability refers to the distribution of normalized histograms (same for Figs. 3 and 4).

*Magnetic and positional symmetry breaking in AFM and PM phases:* The symmetry-broken DFT calculations (Figure 2) show that the AFM phase of $NaOsO_3$ is insulating whereas the PM phase is metallic, in agreement with experiments [1,2]. To establish if the metallicity of PM $NaOsO_3$ is due to the group type being $A^{1+}B^{5+}(O^{2-})_3$ rather than $A^{3+}B^{3+}(O^{2-})_3$, we calculated the former group type but replaced the 5*d* element Os by the *isovalent* 3*d* element Fe using the same orthorhombic crystallographic structure (*Pnma*) with reoptimized lattice constants. We found that both AFM and PM phases of $NaFeO_3$ are insulating (Fig. 2), suggesting the requirement of Fe substituting Os to open a PM gap.

The density of states (DOS) in Fig. 2 indicates three important observations: (i) Without involving strong correlation, one can obtain insulating phases in both AFM and PM phases of $NaFeO_3$ by using symmetry-broken mean-field-like DFT. This is different than Mott mechanism that relies on the effect of strong correlation. (ii) The PM phases that do not have LRO can be either insulating or metallic, depending on 3*d* or 5*d* B-site elements of Fe and Os, respectively. This is beyond Slater mechanism that emphasizes the effect of magnetic LRO for AFM gapping. (iii) In transitioning from the AFM to PM phases, the band gap of $NaFeO_3$ unusually increases from 0.17 eV to 0.29 eV (see Table I and Figure S1 in Supplemental Material) whereas the gap of $NaOsO_3$ decreases from 0.22 eV to zero, reflecting that there could be competing factor to determine how the band gap changes relative to the AFM gap. This is different from the understanding of weak symmetry breaking in persistent metals such as $SrVO_3$. We will illustrate later the two competing factors being magnetic and positional symmetry breaking modes.

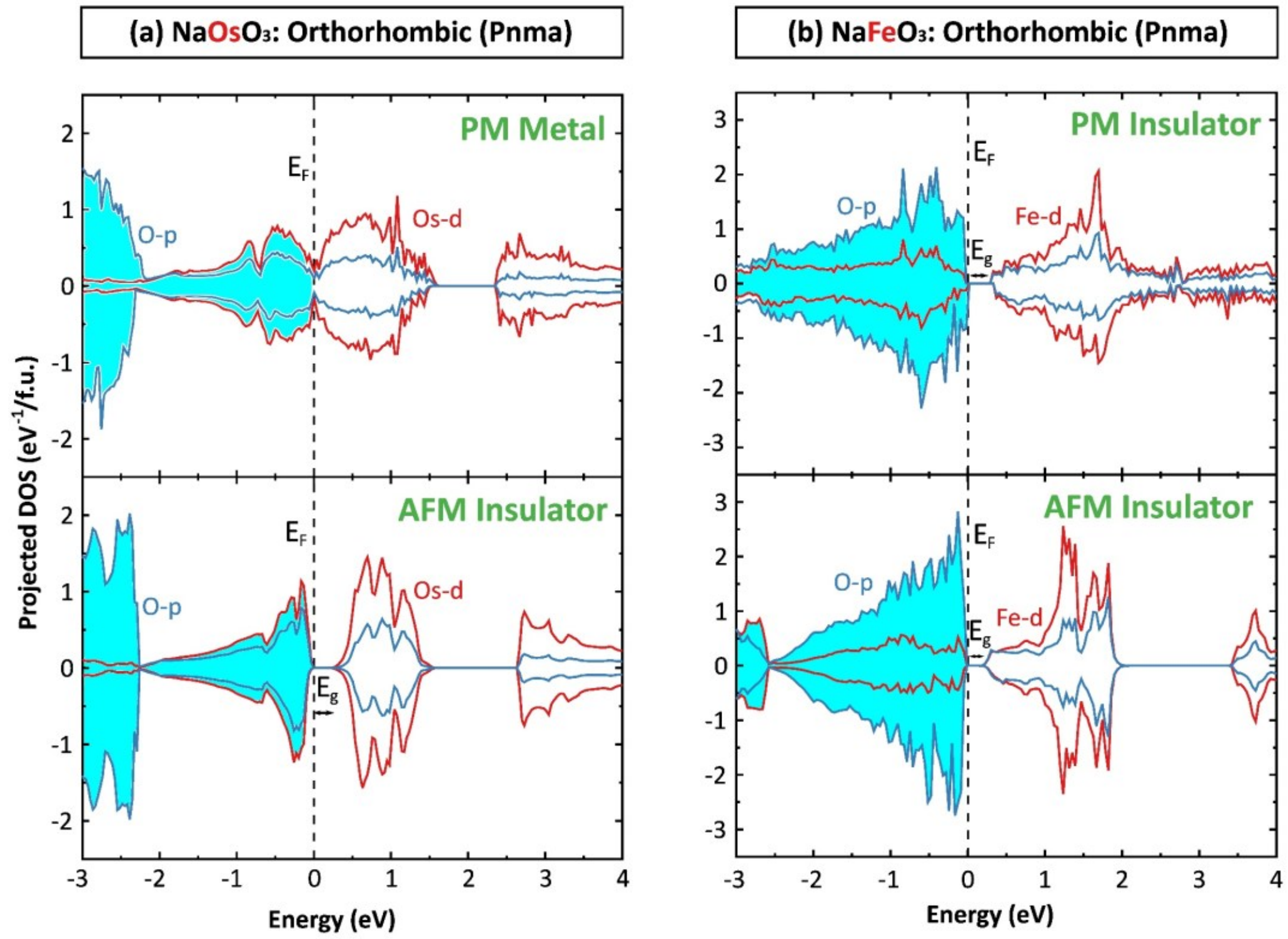

Figure 2. DFT-calculated projected density of states of PM and AFM phases of (a) $NaOsO_3$ and (b) $NaFeO_3$ that have the orthorhombic crystallographic structures (Pnma). Both AFM phases are insulating, but the PM $NaOsO_3$ is metallic while the PM $NaFeO_3$ is insulating.

Figure 3 shows the distribution of local magnetic moments in PM and AFM phases of $NaOsO_3$ and $NaFeO_3$. Although the net total magnetization is zero, there exists non-zero local magnetic moments in both PM and AFM phases. The distribution of magnetic moments in PM phases cannot be mimicked by the nonmagnetic phases that are often assumed in many works [11–14,28,29] on transition-metal oxides in the view of strong correlation as the exclusive factor.

*The broadening of the distribution of magnetic moments in PM phases reflects the effect of spin disorder:* The magnitudes of local magnetic moments in AFM $NaOsO_3$ and AFM $NaFeO_3$ are 1.49 $\mu_B$ and 2.37 $\mu_B$, respectively. The present DFT calculated local moments (such as for Os) are somewhat larger than the experimental value (~1 $\mu_B$ in AFM $NaOsO_3$ [26]). This overestimation could be due to the more spread-out 5*d* orbital in Os by the SCAN functional, although previous DFT calculations using LDA functional *with an adjustable U value* [25] reproduced the ~1 $\mu_B$ ordered moment. Furthermore, the earlier work treated the metallic PM phase (local moments should be nonzero while the total moment is zero) as a nonmagnetic state (local moments are taken to be zero which automatically gives a metal) [25]. In contrast, the present work has no adjustable parameters such as $U$, using instead the SCAN functional and explicitly obtains disordered local moments in the PM phase. The qualitative overestimation of the AFM moment does not affect the following discussion of PM local moments. This magnitude of local magnetic moments in AFM phases reflects the $d^3$ open-shell configuration of highest-spin states that occupies three $t_{2g}$ states in both $NaOsO_3$ and $NaFeO_3$ compounds. The PM phase that has spin disorder inherits the magnetic patterns of the AFM phase, usually broadening the distribution of local magnetic moments of the AFM phase. However, as shown in Fig. 3, the broadening of moments distribution in PM $NaFeO_3$ is negligible and close to the moments in the AFM phase with a small decreased average moment magnitude, but in PM $NaOsO_3$ the broadening relative to moments distribution in the AFM phase becomes rather obvious and generate the magnetic distribution of small and large moment magnitudes of 0.42-0.76 $\mu_B$ and 1.29-1.60 $\mu_B$, respectively. Experimental measurements by resonant inelastic X-ray scattering (RIXS) revealed strongly damped, short-ranged spin fluctuations in the metallic PM phase [30]. These observations are consistent with the DFT-obtained local-moment distribution resolved here (Fig. 3), although RIXS probes dynamical spin correlations and does not directly determine the site-resolved moment magnitudes in our spin-SQS PM configurations.

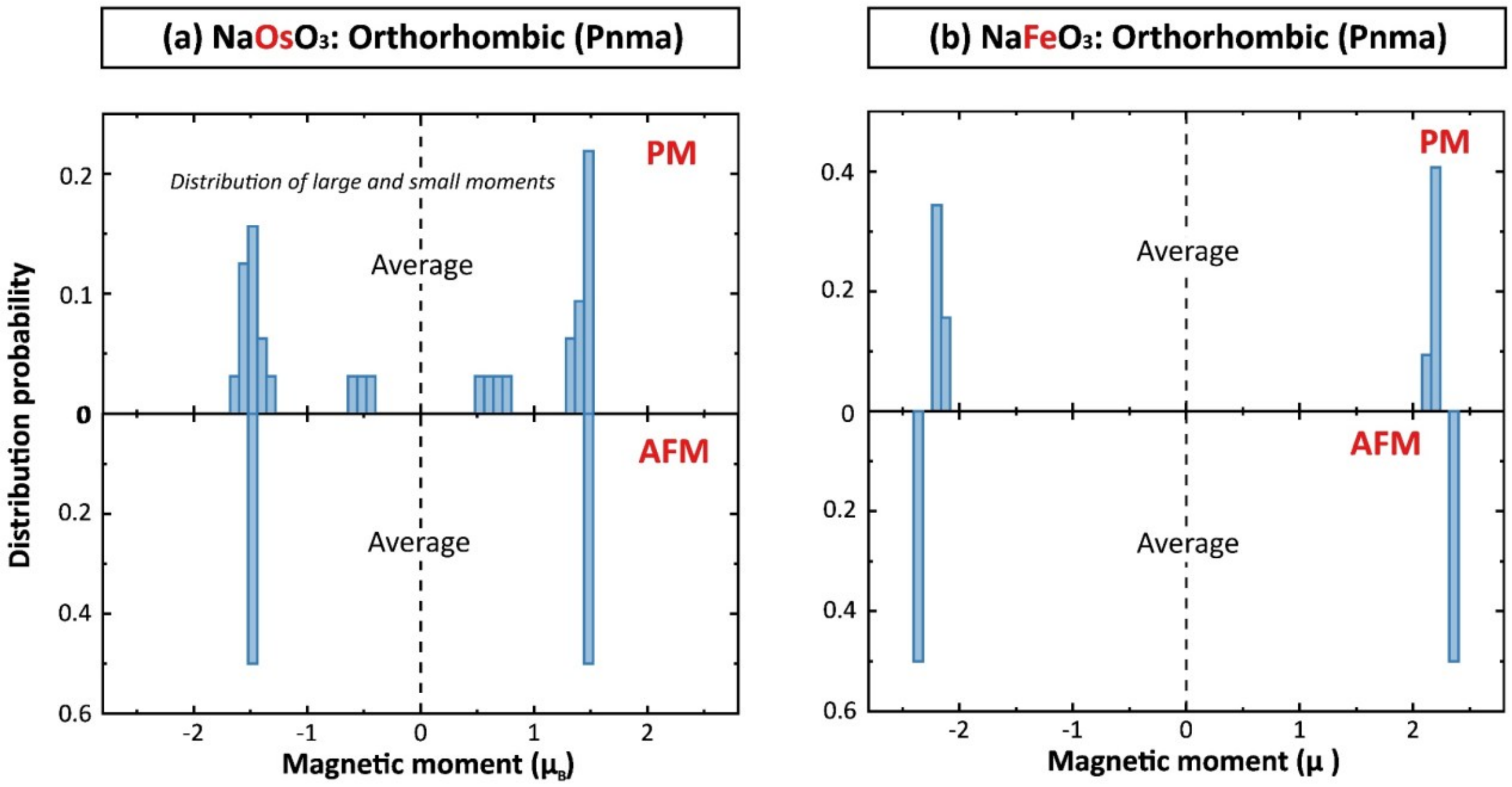

Figure 3. DFT-calculated normalized distribution probability of local magnetic moments in PM and AFM phases of (a) $NaOsO_3$ and (b) $NaFeO_3$ that have the orthorhombic crystallographic structures (Pnma). The vertical dashed lines show the average of all local magnetic moments being zero. All the bin widths of local magnetic moments are set to be 0.08 $\mu_B$.

*Spin-disorder-induced coexistence of large and small local magnetic moments results in metallic PM $NaOsO_3$:* Fig. 3(b) shows that in *PM $NaFeO_3$* there is but a small broadening of moments distribution, indicating a small effect of spin disorder. However, in *PM $NaOsO_3$* [Fig. 3(a)] there is a significant broadening of moments distribution showing the important effect of spin disorder that creates a coexistence of large and small local magnetic moments. Remarkably, this spin-disorder-induced magnetic moments distribution in PM $NaOsO_3$ closes the band gap, leading to a metallic phase.

This can be understood as follows: The moments distribution of $d^3$ configuration in PM $NaOsO_3$ implies that there exists both highest- and lowest-spin states, corresponding to large and small local moments as shown in the top panel of Fig. 3(a). Although the highest-spin state $t_{2g,\uparrow}^3$ is fully occupied, the lowest-spin state $t_{2g,\uparrow}^2 t_{2g,\uparrow}^1$ is partially occupied that leads to the metallic phase. As a comparison, for AFM $NaOsO_3$ and AFM and PM $NaFeO_3$, however, no lowest-spin states are observed (Fig. 3). Thus, the highest-spin states are fully occupied. $YNiO_3$ provides an instructive comparison with the current broader symmetry-breaking mechanism: the competition between a distribution of magnetic moments and a distribution of positional symmetry breaking can select either a metallic or insulating PM state [3,31–34]. For AFM and PM $YNiO_3$, before electron transfer, two electrons instead of three occupying the minority-$t_{2g}$ orbital in the highest-spin state, and one electron occupying the $e_g$ orbital in the lowest-spin state, lead to open shells; after electron transfer, both the highest- and lowest-spin states are fully occupied, leading to closed-shells; thus, the energy reduction driven magnetic disproportionation opens the band gap. As a comparison with no charge transfer in PM $NaOsO_3$, however, there is no electron transfer between different $OsO_6$ octahedra, and hence no magnetic disproportionation, preserving the partially occupied metallic states. Details are given in Section C of Supplemental Material.

*Spin disorder does not explain the increased band gap in PM $NaFeO_3$:* As spin polarization opens the band gap in AFM phases of both $NaOsO_3$ and $NaFeO_3$, we set the AFM band gap as the reference points. Transitioning from AFM to PM phases, the $NaOsO_3$ band gap is closed whereas the $NaFeO_3$ band gap increases. Both cases are unusual compared to most transition-metal oxides in Group I of Table I, in which both AFM and PM phases are insulating with the PM band gaps smaller than AFM gaps. Given that the strong spin-disorder-induced moments distribution in PM $NaOsO_3$ results in the metallic phase, it is still unclear what factor increases the band gap of PM $NaFeO_3$, which will be discussed later.

*Allowing for polymorphous network can increase or even open the band gap of closed-shell nonmagnetic quantum materials:* In closed-shell nonmagnetic quantum materials such as cubic $BaTiO_3$ [35,36], cubic $CsPbI_3$ [20], and $\delta$-$Bi_2O_3$ [19], the polymorphous configuration with different displacements of chemically identical atoms can increase the band gap (in $BaTiO_3$ and $CsPbI_3$) and even open the band gap from a zero (metallic) value (in $\delta$-$Bi_2O_3$) by increasing the asymmetry of *p-d* coupling between cations and anions that dominates the band-edge states. This polymorphism in closed-shell nonmagnetic compounds inspires us to examine its effect in open-shell PM $NaOsO_3$.

*Off-center atomic displacements in PM $NaOsO_3$ and $NaFeO_3$:* The polymorphism can be characterized by the off-center B-site atomic displacements, defined as the distribution of distances between B-site atoms and the centers of six surrounding O atoms in each $BO_6$ octahedron [18]. Figure 4 shows that the off-center B-site atomic displacements in PM $NaOsO_3$ is negligible (< 0.05 Å,) which is an order of magnitude smaller than that in PM $NaFeO_3$ being 0.18-0.19 Å. This large magnitude of off-center displacements of orthorhombic PM $NaFeO_3$ is comparable to or even larger than that of cubic $BaTiO_3$ (0.07-0.09 Å), $CaTiO_3$ (0.01-0.12 Å), and $SrTiO_3$ (0.06-0.08 Å) [18], suggesting an important role in producing the gap in PM $NaFeO_3$.

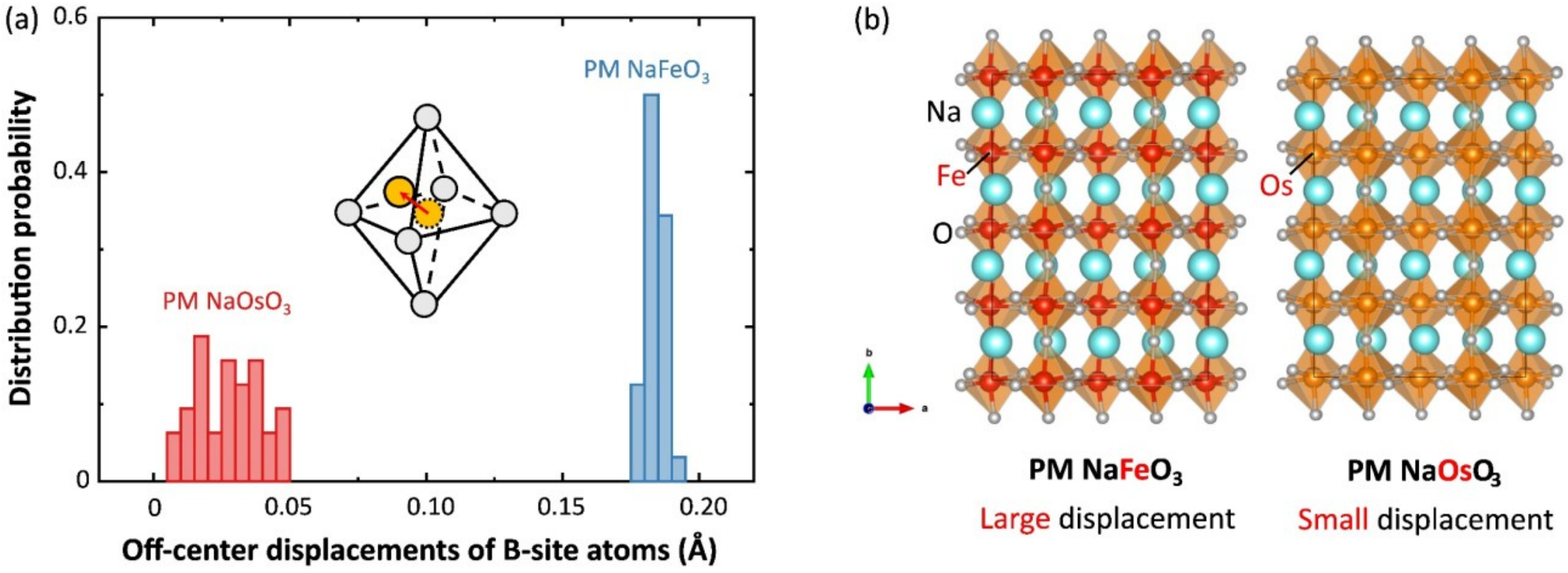

Figure 4. (a) Normalized distribution probability of the octahedral off-center displacements of B-site atoms in PM phases of $NaOsO_3$ and $NaFeO_3$ with (b) orthorhombic crystallographic structures (Pnma). Both PM structures are obtained by DFT relaxation independently. The bin widths of atomic displacements in (a) are set to be 0.005 Å.

*Evolution of the band gap of PM $NaFeO_3$ by artificially tuning the off-center atomic displacements:* Move forward to examine the effect of polymorphism on the band gap of PM $NaFeO_3$. First, we build the no-displacement structure (Structure 1) based on the relaxed PM $NaFeO_3$ insulating structure (Structure 2) by artificially shifting the Fe atoms to the center of each $FeO_6$ octahedron with all other atoms fixed. The calculation of this no-displacement PM phase gives a metallic [see top panel of Figure 5(a)]. Then we use these two structures as end points to construct a series of PM structures by the linear-interpolated method. We obtain the band gaps by calculating these structures as shown in Fig. 5(b), where the two end point structures are marked by red points. As the average off-center atomic displacements increase from zero (Ratio=0) to 0.110 Å (Ratio=0.6), the PM phase stays metallic; When the average displacements further increase from 0.110 Å (Ratio=0.6) to 0.183 Å (Ratio=1), the band gap appears in the PM phase. Fig. 5(a) shows the evolution of DOS at different average off-center Fe displacements. One can see that as the average off-center displacement increases, the conduction band is gradually shifted up till lifting the orbital degeneracy and producing the band gap. Thus, the band gap opening is a result of off-center atomic displacements.

*Spin-disorder-induced local magnetic moments distribution competing with displacements to govern metallic vs insulating PM phase:* In PM $NaFeO_3$, positional polymorphism beats spin disorder and produces a band gap. This can be seen by the large off-center Fe displacements (Figs. 4 and 5) and distribution of local moments without the distribution of large and small local magnetic moments. In PM $NaOsO_3$, however, the metallic phase remains in artificially tuned large off-center Os displacements even reaching 0.346 Å (Ratio=12.8), which we confirm by DFT calculations of the PM structures at different average off-center Os displacements that are constructed by the same linear-interpolated method. The reason is that spin-disorder induced magnetic moments distribution exists in all PM $NaOsO_3$ structures even when the off-center Os displacements are artificially tuned to be large. This is explained by the mixed lowest-spin states that are partially occupied in the $d^3$ occupation (see Fig. S2 in the Supplemental Material). We conclude that spin-disorder-induced magnetic moments distribution wins the competition with polymorphism in PM $NaOsO_3$.

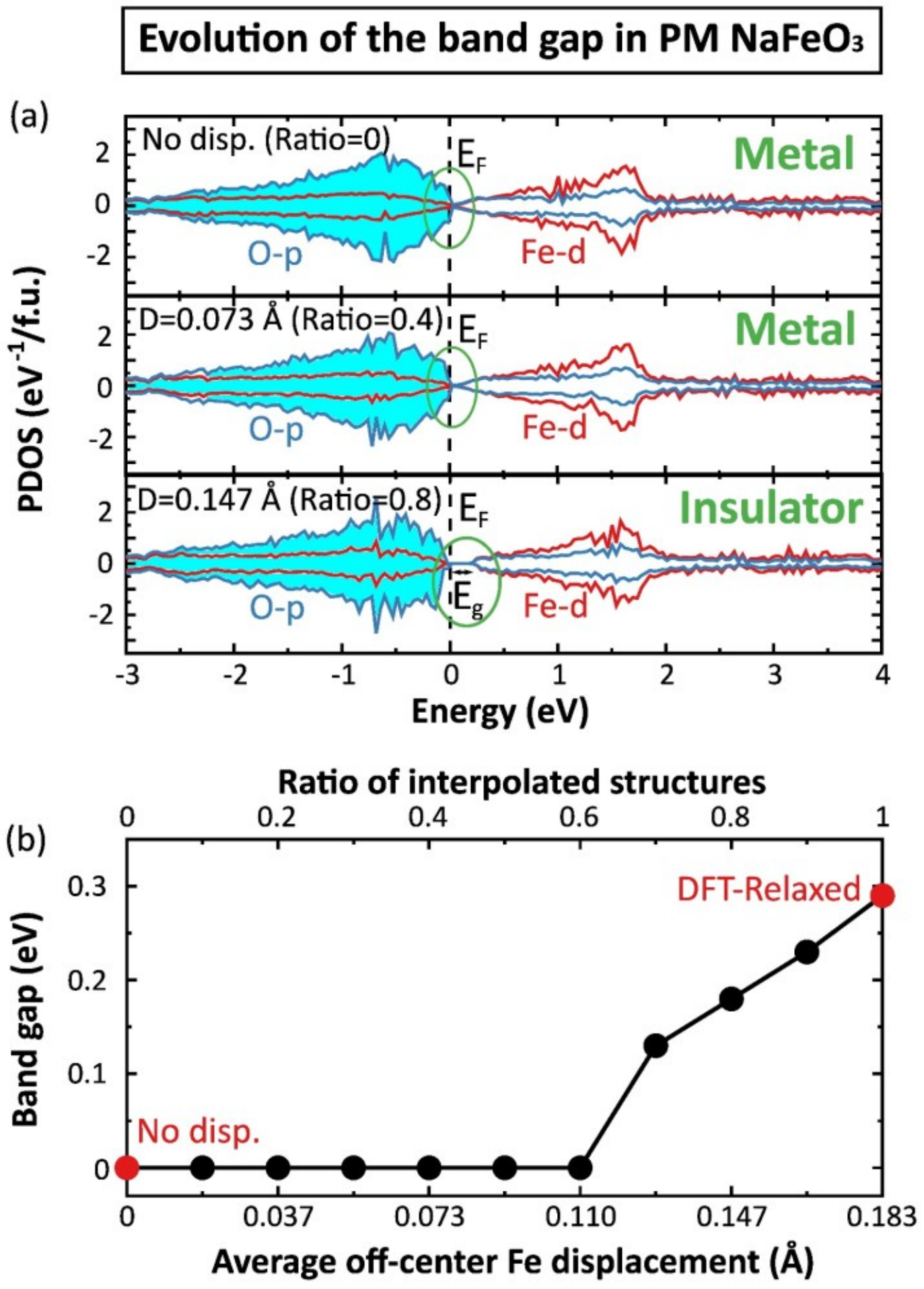

Figure 5. Evolution of the band gap in PM $NaFeO_3$: (a) Density of states and (b) PM band gap as a function of average off-center Fe displacement at different ratios of linear interpolated structures. The two red points in (b) represent the two ending points of structures that are used for the basis of linear interpolation.

The main discoveries in this work are summarized as follows: (a) Both the AFM insulating phase and the PM metallic phase in $NaOsO_3$ can be well described by energy-lowering symmetry-broken DFT approach without strong correlation. (b) Both AFM and PM phases in $NaFeO_3$ are predicted to be insulating, as opposed to the PM metallic phase in $NaOsO_3$. (c) In PM $NaOsO_3$, a strong spin disorder with a distribution of large and small local magnetic moments and negligible Os off-center atomic displacements are predicted, explaining the metallic phase. (d) In PM $NaFeO_3$, a weak spin disorder with strong polymorphous Fe off-center atomic displacements is predicted, leading to the insulating phase. (e) The PM metallic $NaOsO_3$ and unusual large band gap in PM $NaFeO_3$ is explained by the competition between spin-disorder-induced magnetic moments distribution and the strength of B-site off-center atomic displacements in the polymorphous configurations. The approach used is predictive—allowing either metallic or insulating PM behavior–depending on the specific electronic and magnetic descriptors of the specific compounds.

## Acknowledgements

This work was supported by the U.S. Department of Energy (DOE), Office of Science, Basic Energy Sciences, Materials Sciences and Engineering Division, under Grant No. DE-SC0010467. This work used resources from the National Energy Research Scientific Computing Center (NERSC), which was supported by the Office of Science of the DOE.

# Supplemental Material to "Spin disorder competing with positional symmetry breaking determines the metallic versus insulating nature of transition-metal oxide paramagnets"

Jia-Xin Xiong[1], Xiuwen Zhang[1], and Alex Zunger[1]

[1]Renewable and Sustainable Energy Institute, University of Colorado, Boulder, Colorado 80309, USA

## Section A: Details of DFT calculation methods

We set our DFT calculation parameters in VASP [1,2] as follows: (i) The cut-off energy for the plane-wave basis is 520 eV; (ii) The criterion for total energy convergence is $10^{-6}$ eV/cell; (iii) The criterion for relaxations is $10^{-2}$ eV/Å; (iv) The $\Gamma$-centered *k*-point mesh is 3×4×4 for primitive cells and 1×2×2 for supercells in relaxations and self-consistent calculations, and doubled in the calculations of density of states. We use the following PAW-PBE pseudopotentials [3–5] of VASP to represent the ion cores and to reduce the number of explicit electrons. The pseudopotential valence electrons of different elements include Na($2p^6 3s^1$), O($2s^2 2p^4$), Fe($3p^6 3d^7 4s^1$), Os($5p^6 6s^2 5d^6$), Ti($3p^6 3d^2 4s^2$), Mn($3p^6 3d^5 4s^2$), Ni($3p^6 3d^8 4s^2$), La($5s^2 5p^6 5d^1 6s^2$), and Y($4s^2 4p^6 4d^1 5s^2$).

## Section B: Comparison of $NaOsO_3$ structures between symmetry-broken DFT and neutron and X-ray diffraction measurements

We compare the symmetry-broken DFT-relaxed AFM and PM structures of orthorhombic $NaOsO_3$ with the neutron and X-ray diffraction measurements by Calder et al. [6] for the AFM and PM phases at 390.6 K and 430.1 K, respectively. The DFT-relaxed lattice constants are a=5.36 Å, b=7.55 Å, c=5.31 Å for the AFM phase, differing by less than 0.8% from the corresponding experimental values of a=5.39 Å, b=7.60 Å, c=5.35 Å. The PM spin-SQS calculations use the same DFT-relaxed lattice constants as the AFM phase with all cell-internal atomic positions allowed to relax. The experimentally refined PM lattice constants at 430.1 K are a=5.39 Å, b=7.61 Å, c=5.36 Å.

Table SI summarizes the mean values and standard deviations of the Os-O bond lengths, Os-Os atomic distances, Os-O-Os bond angles, and O-Os-O bond angles obtained from the symmetry-broken DFT structures and those from the experimentally refined structures. As discussed in the main text, the DFT-relaxed Os-O bond lengths and Os-Os atomic distances closely reproduce the experimental values. The main residual difference occurs in the Os-O-Os bond angles, which are underestimated by ~3° in the DFT-relaxed structures. To assess the origin of this difference, we additionally performed constrained relaxations using the experimental lattice constants while allowing the cell-internal atomic positions to relax. The mean Os-Os atomic distance then becomes 3.80±0.00 Å in both AFM and PM phases, essentially identical to the experiment. The mean Os-O-Os bond angle increases from 151.96±0.26° to 153.60±0.25° in the AFM phase and from 151.44±1.90° to 152.93±1.62° in the PM phase, approaching the experimental values of 155.68±0.65° and 155.83±0.39°, respectively. The mean Os-O bond length almost does not change. These tests show that the difference in Os-O-Os bond angles is largely associated with the small difference in lattice constants between DFT-relaxation and experiment.

Table SI. Mean value of structural parameters of Os-O bond lengths, Os-Os atomic distances, and Os-O-Os and O-Os-O bond angles between theoretical symmetry-broken DFT-relaxation (the present work) and neutron and X-ray scattering experiment [6] for AFM and PM phases of orthorhombic $NaOsO_3$. The experimental measurements of AFM and PM phases are performed at 390.6 K and 430.1 K, respectively.

| Structural quantity (Mean value) | AFM | | PM | |
|---|---|---|---|---|
| | Theory | Experiment | Theory | Experiment |
| Os-O bond length (Å) | 1.95±0.00 | 1.94±0.00 | 1.95±0.02 | 1.94±0.00 |
| Os-Os atomic distance (Å) | 3.77±0.00 | 3.80±0.00 | 3.77±0.01 | 3.80±0.00 |
| Os-O-Os bond angle (°) | 151.96±0.26 | 155.68±0.65 | 151.44±1.90 | 155.83±0.39 |
| O-Os-O bond angle (°) | 180.00±0.00 | 180.00±0.00 | 178.25±0.98 | 180.00±0.00 |

Table SII addresses a different aspect of the structural comparison by listing the distribution/range of local Os-O bond lengths, Os-Os atomic distances, and Os-O-Os and O-Os-O bond angles in the fully-relaxed DFT configurations. The AFM structure exhibits only narrow variations among the mean values of all the structural parameters, reflecting approximately equivalent local environments. In contrast, the spin-SQS PM structure develops pronounced local distributions: The Os-O bond lengths span 1.92—1.99 Å, Os-Os atomic distances 3.73—3.81 Å, Os-O-Os bond angles 148.19—155.91°, and Os-Os-O bond angles 175.90—179.85°. The experimental values in Table SII have a different meaning from the theoretical ones. Conventional Bragg diffraction determines a periodic crystallographic structure obtained by spatial and temporal averaging. Thus, the quantities such as bond lengths and bond angles correspond to crystallographic inequivalent bonds and angles in the refined average *Pnma* structure, rather than the distribution of local environments represented by the polymorphous spin-SQS PM configuration. The preservation of average *Pnma* symmetry above the Néel temperature therefore does not exclude the presence of locally inequivalent atomic environments.

Table SII. Distribution of structural parameters of Os-O bond lengths, Os-Os atomic distances, and Os-O-Os and O-Os-O bond angles between theoretical symmetry-broken DFT-relaxation (the present work) and neutron and X-ray scattering experiment [6] for AFM and PM phases of orthorhombic $NaOsO_3$. The experimental measurements of AFM and PM phases are performed at 390.6 K and 430.1 K, respectively.

| Structural quantity (Distribution) | AFM | | PM | |
|---|---|---|---|---|
| | Theory | Experiment | Theory | Experiment |
| Os-O bond length (Å) | 1.94—1.95 | 1.94—1.95 | 1.92—1.99 | 1.94—1.95 |
| Os-Os atomic distance (Å) | 3.77 | 3.80 | 3.73—3.81 | 3.80 |
| Os-O-Os bond angle (°) | 151.78, 152.33 | 155.22, 156.59 | 148.19—155.91 | 155.56, 156.39 |
| O-Os-O bond angle (°) | 180.00 | 180.00 | 175.90—179.85 | 180.00 |

Figure 1 in the main text visualizes these local structural distributions in the DFT-relaxed AFM and PM configurations, including distributions of Os-O bond lengths and Os-Os atomic distances [Fig. 1(a)] and distributions of Os-O-Os and O-Os-O bond angles [Fig. 1(b)]. The AFM distributions remain narrow, whereas all the four quantities broaden in the spin-SQS PM configuration. This broadening directly illustrates the inequivalent local positional environments in the spin-SQS PM configuration, which cannot be captured by a single set of average crystallographic coordinates as measured in the neutron and X-ray scattering experiment [6].

## Section C: DFT-calculated band gaps of Group I transition-metal compounds

Group I transition-metal compounds, as shown in Table I in the main text, have low-temperature (low-T) ground AFM phases and high-temperature (high-T) PM phases that are both insulating. We did the spin-polarized DFT calculations for AFM and PM phases of Group I compounds using the SCAN functional without the on-site potential term *U*. Primitive cells are used for the AFM phases, and supercells that contain non-zero local magnetic moments produced by spin special-quasirandom structure (spin-SQS) [7] are adopted for the PM phases. The calculated results are shown in Figure S1.

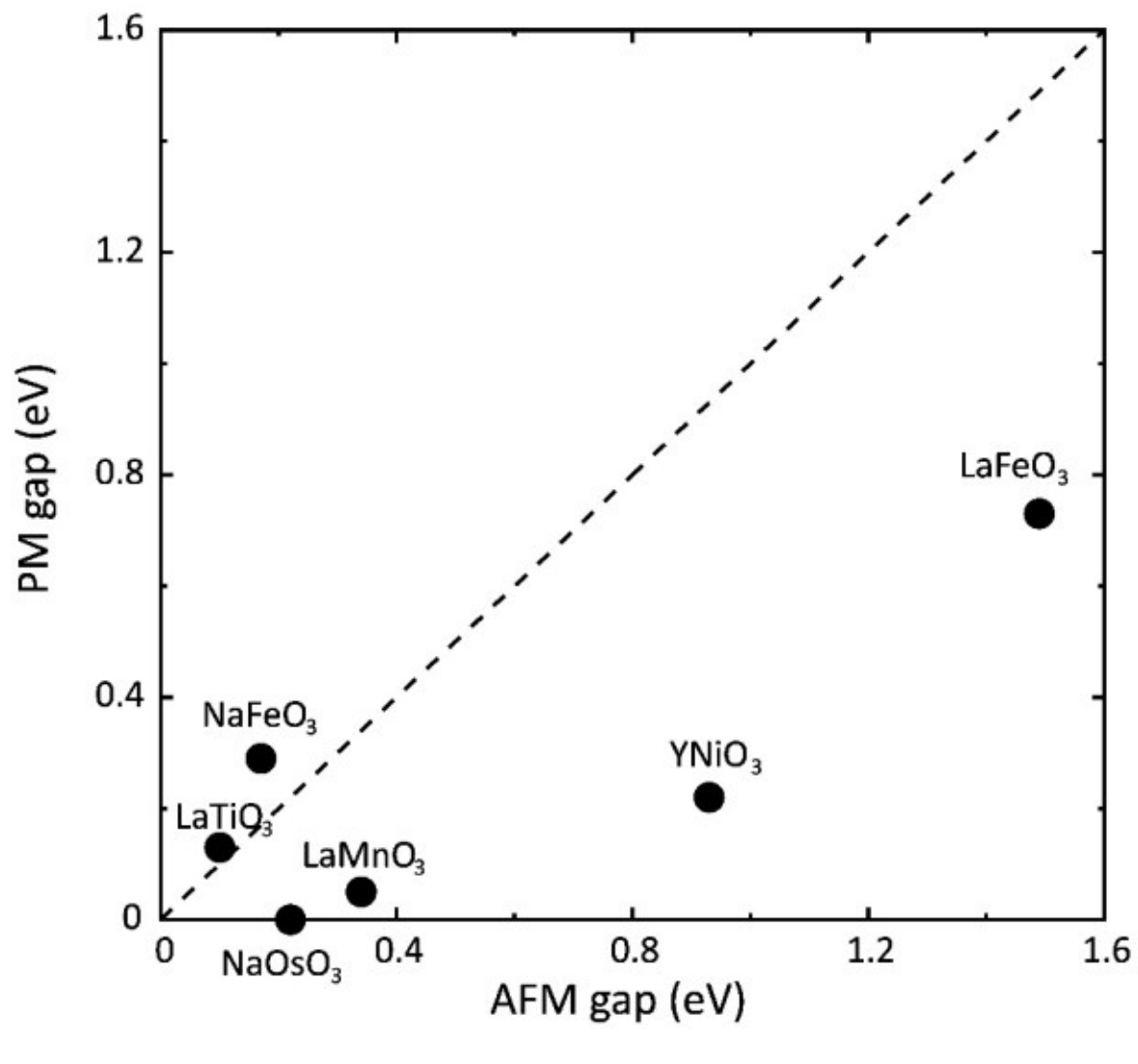

Figure S1. DFT-calculated AFM and PM gaps of all compounds listed in Table I in the main text by SCAN.

## Section C: Distribution of large and small local magnetic moments in transition-metal compounds PM $NaOsO_3$ and $YNiO_3$

Both the $d^7$ compound $YNiO_3$ in its AFM and PM phases [8] and the $d^3$ compound $NaOsO_3$ in its PM phase have a distribution of large and small local magnetic moments. This distribution of local moments indicates the coexistence of highest- and lowest-spin states in both compounds, corresponding to the large and small local moments, respectively. However, the small local moments in $YNiO_3$ are around zero (Fig. 5 in Ref. [8]), whereas those in $NaOsO_3$ are non-zero (Fig. 3 in the main text). This is explained in Figure S2: Electron transfer only exists in $YNiO_3$ and leads to magnetic disproportionation of $d^8$ and $d^6$ configurations. In PM $NaOsO_3$, however, there is no magnetic disproportionation. The magnetic disproportionation in $YNiO_3$ makes its highest- and lowest-spin states both fully occupied and hence opens the band gap. In PM $NaOsO_3$, however, the lowest-spin state is partially occupied though its highest-spin state is fully occupied, leading to the metallic phase.

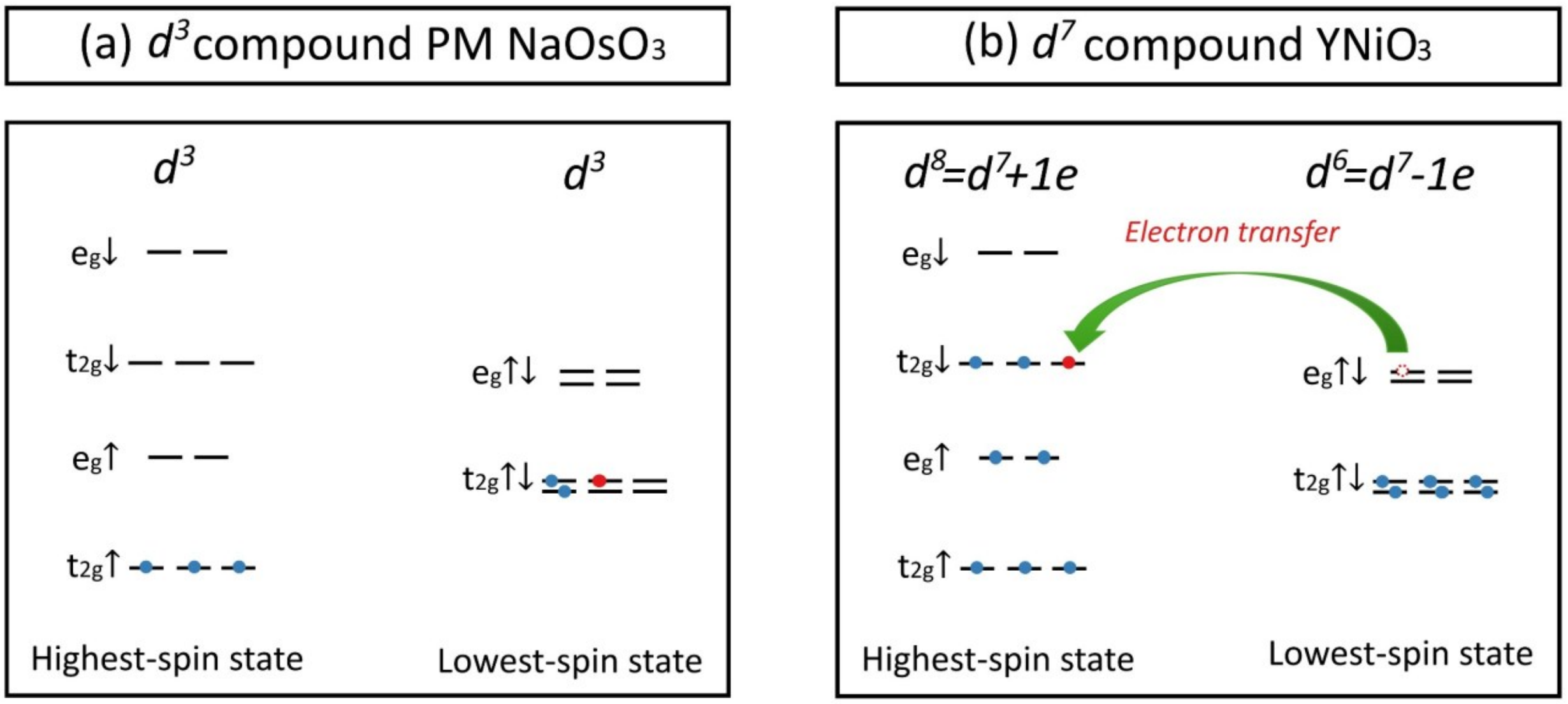

Figure S2. Schematic diagram of occupied energy levels with large and small local magnetic moments in (a) $d^3$ compound PM $NaOsO_3$ and (b) $d^7$ compound AFM and PM $YNiO_3$. There is electron transfer from lowest- to highest-spin state only in (b) $YNiO_3$, leading to the magnetic disproportionation of $d^8$ and $d^6$ motif distributions. The up and down arrows for the energy levels represent majority- and minority-spin states, respectively.

## Supplemental Reference